\documentstyle[a4wide,epsf,11pt,titlepage]{article}

\newcounter{nref}
\newcommand{\bbib}{%
  \renewcommand{\refname}{\large\bf References}%
  \begin{thebibliography}{99}%
  \setcounter{enumiv}{\thenref}}
\newcommand{\ebib}{%
  \end{thebibliography}%
  \setcounter{nref}{\arabic{enumiv}}}
\newcommand{\head}[3]{%
  \setcounter{nref}{0}%
  \thispagestyle{empty}%
  \section*{\LARGE\bf #1}%
  \stepcounter{section}%
  \addcontentsline{toc}{section}{#1}%
  \large\itshape%
  #2\\\vspace{0.1pt}\\%
  #3%
  \normalsize\upshape%
  \bigskip}

\begin{document}


\head{The Importance of Neutrino Opacities for the Accretion Luminosity
in Spherically Symmetric Supernova Models}
     {M.\ Liebend\"orfer$^{1,2}$, O.~E.~B.\ Messer$^{1,2,3}$, A.\ Mezzacappa$^{1}$, W.~R.\ Hix$^{1,2,3}$, F.-K. Thielemann$^{4}$, K. Langanke$^{5}$}
     {$^1$ Physics Division, Oak Ridge National Laboratory, Oak Ridge, TN\-37831 \\
      $^2$ Department of Physics and Astronomy, University of Tennessee, Knoxville, TN\-37996 \\
      $^3$ Joint Institute for Heavy Ion Research, Oak Ridge National 
                Laboratory, Oak Ridge, TN\-37831 \\
      $^4$ Department of Physics and Astronomy, University of Basel, CH\-4056 Basel \\
      $^5$ Institut for Fysik og Astronomi, \r{A}rhus Universitet, DK\-8000 \r{A}rhus C}

\subsection*{General Relativistic Boltzmann Neutrino Transport}
Spherically symmetric simulations of stellar core collapse and postbounce
evolution do not lead to explosions if they are based on ``standard''
\cite{liebend.Bruenn_85,liebend.Lattimer_Swesty_91}
nuclear and weak interaction microphysics. This widely anticipated
statement achieved a completely new level of justification with
the technically complete and self-consistent treatment of neutrino
transport in the highly dynamical evolution after core collapse and
bounce \cite{liebend.Rampp_Janka_00,liebend.Mezzacappa_et_al_01}
in general relativity
\cite{liebend.Bruenn_DeNisco_Mezzacappa_01,liebend.Liebendoerfer_et_al_01}.
Multidimensional models for core collapse supernovae, which allow
non stratified hydrodynamics at the cost of reliable neutrino transport,
present the neglect of convection as a plausible cause for the failure
to reproduce supernova explosions in spherical symmetry (e.g. see Ref.
\cite{liebend.Janka_Kifonidis_Rampp_01} and references therein).
In our spherically symmetric models, neutrino
absorption adds only weakly to the entropy of the shock-dissociated
material after shock stagnation because of high infall velocities.
Nevertheless, spherically symmetric models provide a field for
further studies: The realistic neutron star structure and the accurate
evolution of angle- and energy-resolved neutrino distribution functions
enable the investigation of the influence of nuclear and weak interaction
physics on the models. Complete implicit radiation hydrodynamics also
provides insight into many tightly coupled processes with strong feedback
in the evolution of a core collapse supernova.

In the following, we compare the postbounce evolution of five
different initial stellar progenitors, simulated with general
relativistic three-flavor Boltzmann neutrino transport. The chosen
progenitors have masses of $13$ M$_{\odot}$ and $20$ M$_{\odot}$
\cite{liebend.Nomoto_Hashimoto_88}; and $15$ M$_{\odot}$, $25$
M$_{\odot}$, and $40$ M$_{\odot}$ \cite{liebend.Woosley_Weaver_95}. We
present in Fig. 1 the shock position for these different models as a
function of time, as solid lines. We find similar trajectories for all
progenitors, with a shock radius maximum around $150$ km. We also plot the
shock position in terms of enclosed mass in Fig. 2b: Before the existence
of a shock during core collapse, we plot the deleptonization-dependent
position of the sonic point where the shock will form after bounce. In a
later stage, the gradient of the shock trajectory reflects the accretion
rate given by the radial density profile of the progenitor star. Also
shown are the electron neutrino luminosities and rms energies in
Fig. 2a. Among the chosen progenitors, which represent the full progenitor mass
range of expected type II events, the most striking feature is
the clear separation into a quantitatively similar evolution up to the
electron neutrino burst and a distinctively different one afterwards.

\begin{figure}[ht]
  \centerline{\epsfxsize=0.9\textwidth\epsffile{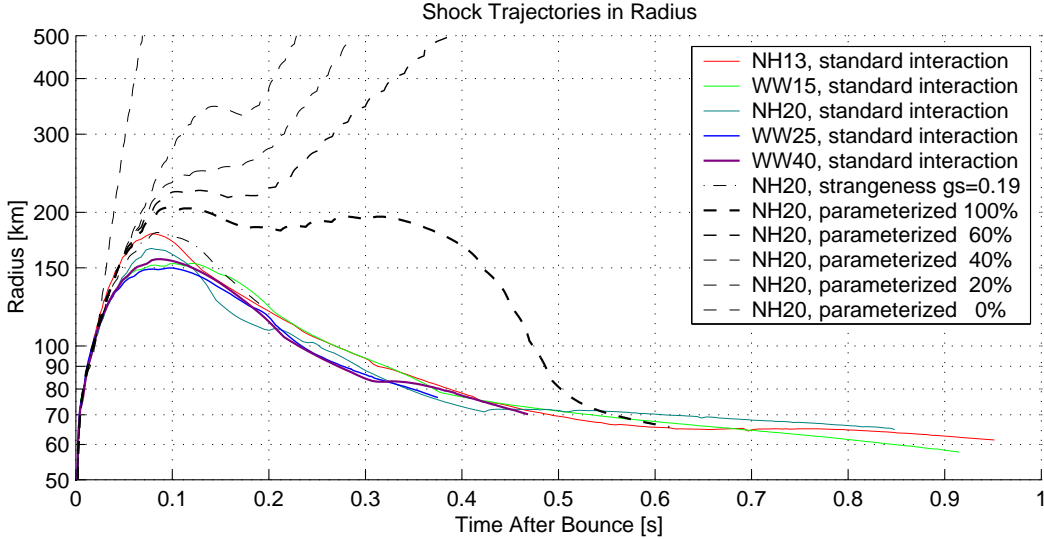}}
  \caption{Shown in solid lines are the shock trajectories with standard input physics. Dash-dotted is the shock trajectory of the 20 solar mass
progenitor with nucleon strangeness content included. The experimental dashed
trajectories (based on transport with lowest angular resolution) show explosions
as a function of parametrized isoenergetic scattering on free nucleons.}
\end{figure}

\subsection*{Feedback in the Deleptonization during Core Collapse} When
the inner core of the progenitors reach the Chandrasekhar mass, they
have comparable density and temperature profiles. The electron fraction
determines the location of the sonic point at the surface of the inner
core, which is causally connected by electron pressure (Fig. 2b). During
collapse, the electron fraction decreases by electron capture on nuclei
and protons until trapped neutrinos block further deleptonization.
A regulation mechanism establishes similar electron fractions in all
models \cite{liebend.Bruenn_85,liebend.Messer_et_al_02}: The small
abundance of free protons, $Y_p$, is given by the Lattimer-Swesty equation
of state as a function of the electron fraction, $Y_e$. The sensitivity
$d\ln (Y_p)/dY_e\sim 30$ is large enough that a small deviation, $\Delta
Y_e=0.01$, from a ``normal'' $Y_e$-evolution changes the proton abundance,
and therewith the number of electron captures on free protons, by a
third. This negative feedback drives the electron fraction back to the
``norm''-evolution whenever electron captures on free protons dominate.
In our simulations, capture on heavy nuclei is suppressed as soon as the
$N=40$ shell is closed \cite{liebend.Bruenn_85}. Thus, the described
self-regulation sets in at this point, leading to very small differences
in the evolution of the different progenitors up to shock breakout.
The extent to which this regulation survives extension of electron
capture rates to heavier nuclei, when these rates are computed
with improved nuclear shell models, remains to be investigated.

\begin{figure}[ht]
  \centerline{\epsfxsize=0.45\textwidth\epsffile{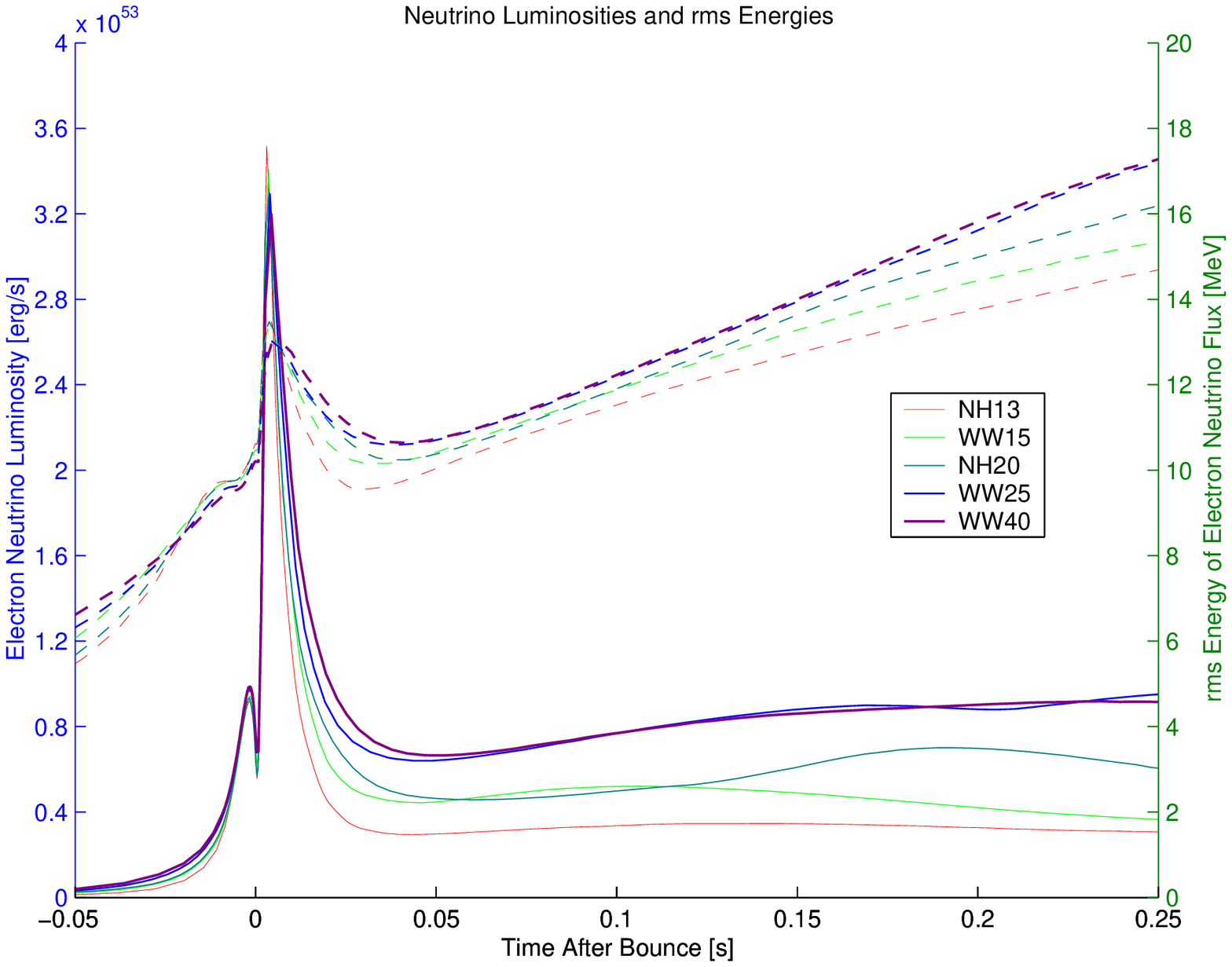}
                      \epsfxsize=0.45\textwidth\epsffile{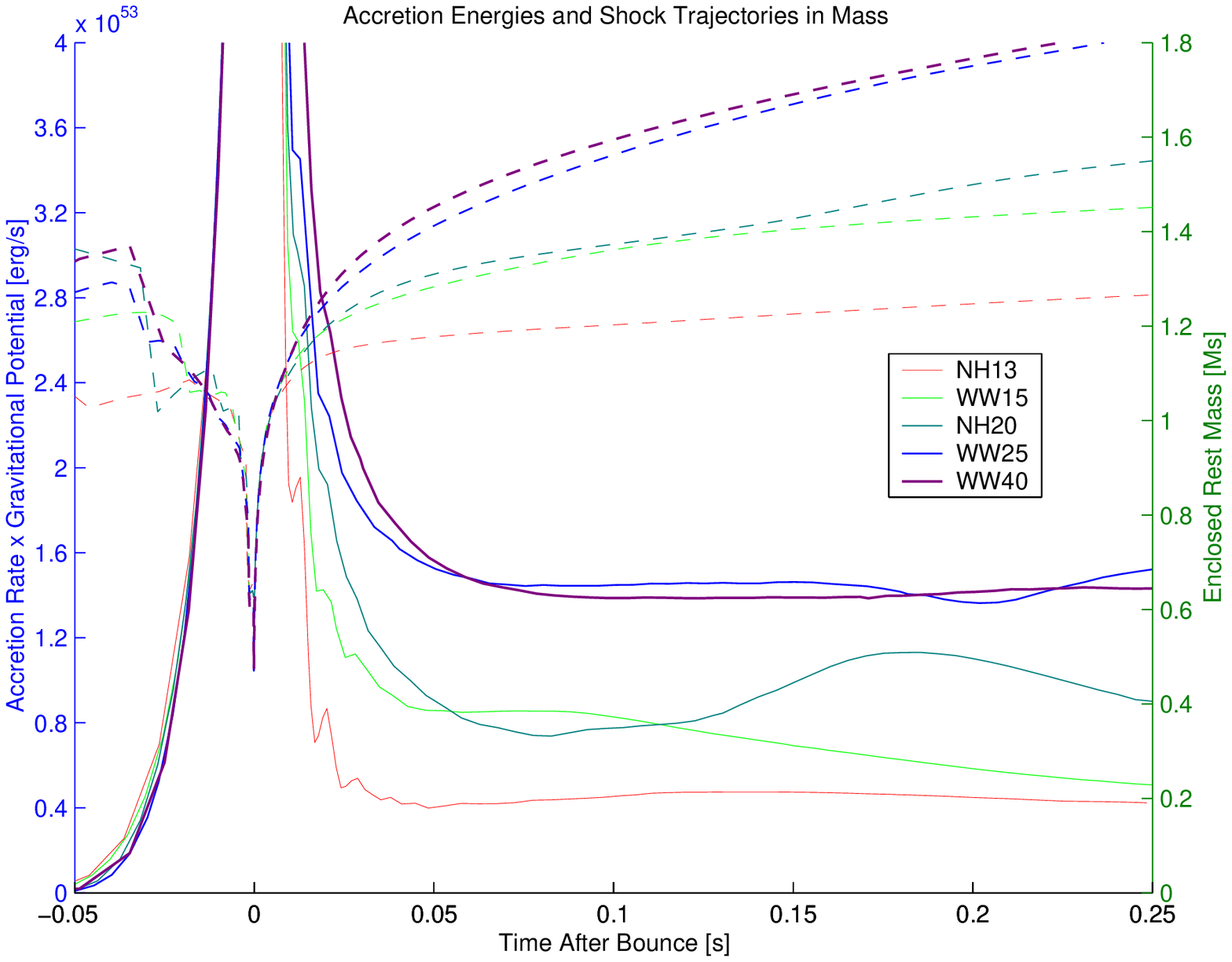}}
  \caption{The left hand side shows the electron neutrino luminosities
(solid) and rms energies (dashed). On the right hand side are the
accretion energy rate (solid) and the sonic point/mass trajectories in
enclosed mass (dashed).} \end{figure}

\subsection*{Accretion Luminosities and Neutrino Opacities}
The accretion luminosity, which dominates over the core diffusion luminosity
soon after shock stall, is produced by
the compression of infalling material at the surface of the PNS. The
available energy can roughly be estimated by the rate of infalling
material, multiplied by the gravitational potential at the neutrino
sphere. The result is an energy deposition rate that qualitatively
relates the modulation in the observable neutrino luminosities to
accretion rates (Fig. 2ab), i.e. to variations in the initial density
profiles of the progenitor stars. A relation of this kind might be
linked to analytical supernova models that present conditions for
shock revival as a function of independent luminosities and accretion
rates \cite{liebend.Burrows_Goshy_93,liebend.Janka_01}. The
main influence of nuclear input physics to the evolution of the shock
position resides in how the high density equation of state determines
the size of the PNS on the one hand, and how the neutrino opacities
at moderate densities ($\sim 10^{13}$ g/cm$^3$) affect the 
energy loss rate by neutrino emission on the other:

Indeed, reduced isoenergetic scattering opacities at moderate densities
have led to explosions in spherical symmetry and suggested a search for
an exploration of the physical limits. However, a physically
justifiable opacity reduction by taking into account the strangeness
content of nucleons (see e.g. \cite{liebend.Kolbe_et_al_92}) did not
produce explosions, even if it had a positive effect in exploratory runs
(Fig. 1). A detailed discussion of the degrees of freedom in the nuclear
input in this regime has been given in Ref. \cite{liebend.Horowitz_02}. A
parameter study with decreased opacities beyond physical limits (Fig. 1)
shows an interesting side effect: In exploding models, we find
electron fractions close to $0.5$ in the inner layers of ejecta
as soon as the electron degeneracy is lifted in the neutrino heated, expanding
material. The electron fraction changes during mass ejection will be
considered in upcoming nucleosynthesis calculations of supernova yields
\cite{liebend.Hauser_et_al_02}.

With standard opacities at moderate densities, the microphysics
above nuclear densities affects the postbounce
evolution more by influencing the size and stability of
the PNS than by influencing the diffusive neutrino flux. A more compact PNS
puts the accretion/heating cycle deeper into the gravitational
potential, resulting in higher infall velocities and luminosities
\cite{liebend.Bruenn_DeNisco_Mezzacappa_01,liebend.Liebendoerfer_et_al_01}.
An explosion-enhancing effect from the deeper gravitational potential
can only be obtained if the PNS contraction is fast enough to produce
very high luminosities before the heating region has adjusted to the
smaller PNS radius, or if the outer layers are kept at a distance
by a yet unidentified mechanism. PNS convection may provide
a more visible consequence of changes in the high density physics
\cite{liebend.Keil_Janka_Mueller_96,liebend.Mezzacappa_et_al_98}.
However, a definitive conclusion regarding the effect of convection
below and around the neutrino sphere awaits a more refined equation
of state and the inclusion of detailed multidimensional neutrino transport.

\subsection*{Acknowledgements}
M.L. is supported by the National Science Foundation under contract 
AST-9877130.
O.E.B.M. is supported by funds from the DoE HENP
Scientific Discovery through Advanced Computing Program. 
A.M. is supported at the Oak Ridge National Laboratory, managed by
UT-Battelle, LLC, for the U.S. Department of Energy under contract
DE-AC05-00OR22725.
W.R.H. is supported by NASA under contract NAG5-8405 and by funds 
from the Joint Institute for Heavy Ion Research and a DoE HENP
PECASE Award.
F.-K.T. is supported in part by the Swiss National Science Foundation 
under contract 20-61822.00.
K. L. is supported in part by the Danish Research Council.
We thank the Institute for Nuclear Theory at the University of
Washington for its hospitality and the Department of Energy for
partial support.
Our simulations were carried out on the National Energy Research
Supercomputer Center Cray SV-1.

\end{document}